\PassOptionsToPackage{unicode}{hyperref}
\PassOptionsToPackage{hyphens}{url}
\PassOptionsToPackage{dvipsnames,svgnames,x11names}{xcolor}
\documentclass[
  12pt]{article}
\usepackage{xcolor}
\usepackage{amsmath,amssymb}
\usepackage{iftex}
\ifPDFTeX
  \usepackage[T1]{fontenc}
  \usepackage[utf8]{inputenc}
  \usepackage{textcomp} 
\else 
  \usepackage{unicode-math} 
  \defaultfontfeatures{Scale=MatchLowercase}
  \defaultfontfeatures[\rmfamily]{Ligatures=TeX,Scale=1}
\fi
\usepackage{lmodern}
\ifPDFTeX\else
\fi
\IfFileExists{upquote.sty}{\usepackage{upquote}}{}
\IfFileExists{microtype.sty}{
  \usepackage[]{microtype}
  \UseMicrotypeSet[protrusion]{basicmath} 
}{}
\makeatletter
\@ifundefined{KOMAClassName}{
  \IfFileExists{parskip.sty}{%
    \usepackage{parskip}
  }{
    \setlength{\parindent}{0pt}
    \setlength{\parskip}{6pt plus 2pt minus 1pt}}
}{
  \KOMAoptions{parskip=half}}
\makeatother
\makeatletter
\ifx\paragraph\undefined\else
  \let\oldparagraph\paragraph
  \renewcommand{\paragraph}{
    \@ifstar
      \xxxParagraphStar
      \xxxParagraphNoStar
  }
  \newcommand{\xxxParagraphStar}[1]{\oldparagraph*{#1}\mbox{}}
  \newcommand{\xxxParagraphNoStar}[1]{\oldparagraph{#1}\mbox{}}
\fi
\ifx\subparagraph\undefined\else
  \let\oldsubparagraph\subparagraph
  \renewcommand{\subparagraph}{
    \@ifstar
      \xxxSubParagraphStar
      \xxxSubParagraphNoStar
  }
  \newcommand{\xxxSubParagraphStar}[1]{\oldsubparagraph*{#1}\mbox{}}
  \newcommand{\xxxSubParagraphNoStar}[1]{\oldsubparagraph{#1}\mbox{}}
\fi
\makeatother

\usepackage{longtable,booktabs,array}
\usepackage{calc} 
\usepackage{etoolbox}
\makeatletter
\patchcmd\longtable{\par}{\if@noskipsec\mbox{}\fi\par}{}{}
\makeatother
\IfFileExists{footnotehyper.sty}{\usepackage{footnotehyper}}{\usepackage{footnote}}
\makesavenoteenv{longtable}
\usepackage{graphicx}
\makeatletter
\newsavebox\pandoc@box
\newcommand*\pandocbounded[1]{
  \sbox\pandoc@box{#1}%
  \Gscale@div\@tempa{\textheight}{\dimexpr\ht\pandoc@box+\dp\pandoc@box\relax}%
  \Gscale@div\@tempb{\linewidth}{\wd\pandoc@box}%
  \ifdim\@tempb\p@<\@tempa\p@\let\@tempa\@tempb\fi
  \ifdim\@tempa\p@<\p@\scalebox{\@tempa}{\usebox\pandoc@box}%
  \else\usebox{\pandoc@box}%
  \fi%
}
\def\fps@figure{htbp}
\makeatother

\providecommand{\tightlist}{%
  \setlength{\itemsep}{0pt}\setlength{\parskip}{0pt}}

\usepackage[]{natbib}
\usepackage{booktabs}
\usepackage{caption}
\usepackage{longtable}
\usepackage{colortbl}
\usepackage{array}
\usepackage{anyfontsize}
\usepackage{multirow}
\usepackage{xcolor}
\makeatletter
\@ifpackageloaded{caption}{}{\usepackage{caption}}
\AtBeginDocument{%
\ifdefined\contentsname
  \renewcommand*\contentsname{Table of contents}
\else
  \newcommand\contentsname{Table of contents}
\fi
\ifdefined\listfigurename
  \renewcommand*\listfigurename{List of Figures}
\else
  \newcommand\listfigurename{List of Figures}
\fi
\ifdefined\listtablename
  \renewcommand*\listtablename{List of Tables}
\else
  \newcommand\listtablename{List of Tables}
\fi
\ifdefined\figurename
  \renewcommand*\figurename{Figure}
\else
  \newcommand\figurename{Figure}
\fi
\ifdefined\tablename
  \renewcommand*\tablename{Table}
\else
  \newcommand\tablename{Table}
\fi
}
\@ifpackageloaded{float}{}{\usepackage{float}}
\floatstyle{ruled}
\@ifundefined{c@chapter}{\newfloat{codelisting}{h}{lop}}{\newfloat{codelisting}{h}{lop}[chapter]}
\floatname{codelisting}{Listing}

\makeatother
\makeatletter
\@ifpackageloaded{caption}{}{\usepackage{caption}}
\@ifpackageloaded{subcaption}{}{\usepackage{subcaption}}
\makeatother
\usepackage{bookmark}
\IfFileExists{xurl.sty}{\usepackage{xurl}}{} 
\hypersetup{
  pdftitle={Active Learning with Bayesian Reasoning: A POGIL-Based Pedagogy in Introductory Statistics},
  pdfauthor={Cheng-Han Yu; Angela Ebeling},
  pdfkeywords={conditional probability, Bayes' theorem, undergraduate
statistics education, Process Oriented Guided Inquiry Learning, Bayesian
bivariate generalized linear model},
  colorlinks=true,
  linkcolor={blue},
  filecolor={Maroon},
  citecolor={Blue},
  urlcolor={Blue},
  pdfcreator={LaTeX via pandoc}}

\begin{document}

\def\spacingset#1{\renewcommand{\baselinestretch}%
{#1}\small\normalsize} \spacingset{1}


\date{June 11, 2026}
\title{\bf Active Learning with Bayesian Reasoning: A POGIL-Based
Pedagogy in Introductory Statistics}
\author{
Cheng-Han Yu\\
Department of Mathematical and Statistical Sciences, Marquette
University\\
and\\Angela Ebeling\\
Department of Biology and Environmental Science, Wisconsin Lutheran
College\\
}
\maketitle

\bigskip
\bigskip
\begin{abstract}
We introduce a Process Oriented Guided Inquiry Learning (POGIL)-style
activity for teaching Bayesian reasoning in introductory statistics
through conditional probability, Bayes' theorem, and belief updating.
The activity is self-contained, uses hand-computable probabilities
organized in two-way tables, and engages students in structured team
roles. We evaluated it in four sections of an undergraduate introductory
statistics course using a quasi-experimental comparison of POGIL-style
and lecture-based instruction for a Bayes' theorem unit. Outcomes
included student performance on Bayes' theorem final exam questions and
satisfaction with instruction. We used a Bayesian bivariate generalized
linear model to compare the two approaches while accounting for major
type, gender, and race. The results indicated similar exam performance
and probabilities of high satisfaction across instructional styles and
demographic groups, with considerable uncertainty and no clear evidence
of meaningful differences. These findings suggest that the POGIL-style
activity performed comparably to lecture-based instruction for this unit
while offering an active and classroom-ready way to introduce Bayesian
reasoning without requiring difficult computation or simulation. We
provide adaptable instructional materials and a reproducible Bayesian
analytic framework for evaluating active learning innovations in
introductory statistics. Our study supports feasible inclusion of
Bayesian reasoning in introductory courses and may help instructors
considering active learning.
\end{abstract}

\noindent%
{\it Keywords:} conditional probability, Bayes' theorem, undergraduate
statistics education, Process Oriented Guided Inquiry Learning, Bayesian
bivariate generalized linear model
\vfill

\newpage
\spacingset{1.9} 

\section{Introduction}\label{sec-intro}

Conditional probability and Bayes' theorem are widely recognized as
conceptually difficult topics in introductory statistics
\citep{Garfield88, Borovcnik91, Kvatinsky02, Cui23}. Students often
struggle to interpret conditional statements and to distinguish
probabilities such as Pr(A \textbar{} B) and Pr(B \textbar{} A), and
instructors report that probability chapters are among the least
engaging and most challenging portions of the course
\citep{Carolyn01, Garfield08}. At the same time, Bayesian reasoning
appears routinely in applications of statistics, machine learning, and
artificial intelligence (AI) \citep{Ghahramani15, Murphy22}. This raises
a question for statistics educators: How can we introduce Bayesian ideas
meaningfully and accessibly at the introductory level?

\begin{figure}

\centering{

\includegraphics[width=0.7\linewidth,height=\textheight,keepaspectratio]{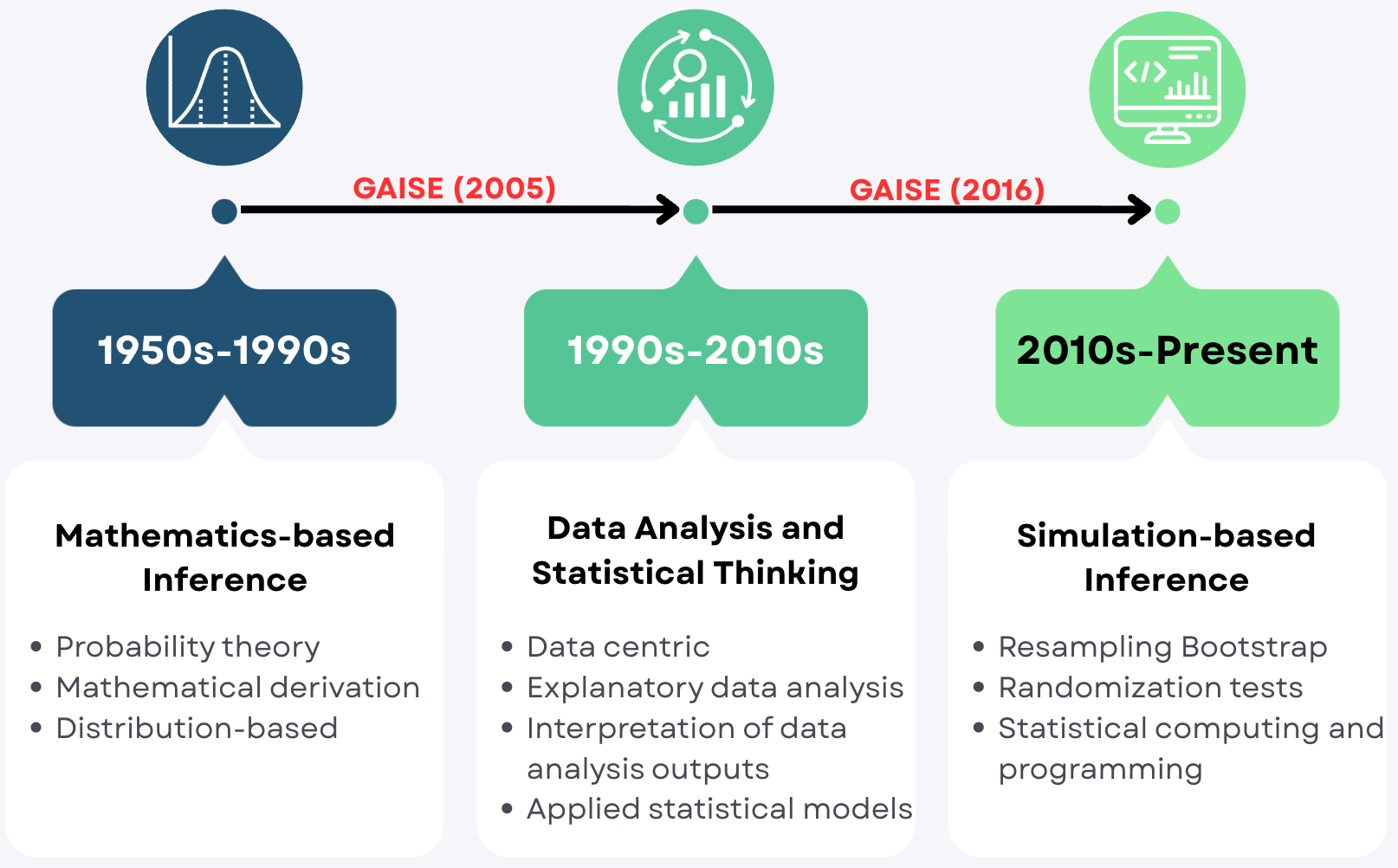}

}

\caption{\label{fig-timeline}Changes in Introductory Statistics. Each
period shows the main focus and topics discussed in introductory
statistics courses. Introductory statistics has moved from
distribution-based and probability-focused to data-centered,
distribution-free, and then simulation-based and computation-based
instruction.}

\end{figure}

Figure~\ref{fig-timeline} summarizes the evolution of the introductory
statistics curriculum from the mid 20th century to the present. Early
texts emphasized a distribution-based and probability-focused
perspective with several chapters on probability and distribution theory
before introducing classical inference \citep{Cobb07, Agresti23}. As
time passed, the curriculum became more data-centered with emphasis on
real data, exploratory data analysis, and statistical thinking
\citep{GAISE05, GAISE16}. Recently, simulation- and computation-based
approaches have become increasingly prominent, especially through
resampling- and randomization-based inference \citep{Tintle16, Lock21}.

The curriculum has evolved in how it motivates and teaches classical
inference, but it typically does not introduce Bayesian inference as an
alternative framework, even though teaching Bayesian inference in
introductory courses has been proposed
\citep{albert01, Cobb07, Baglin09, Hoegh20}.
\begingroup\color{black}There are complications to consider when trying
to incorporate both frequentist and Bayesian approaches into an
introductory statistics course -- these include the need for students to
learn both sampling distributions and conditional probabilities (both
difficult concepts), the confusion on the part of students related to
interpretation of the results of frequentist and Bayesian approaches,
and the relative dearth of Bayesian approaches in the literature of the
disciplines students may be studying \citep{albert95, moore97}.
Additionally, few texts are available for teaching Bayesian inference at
a level where the only prerequisite is college algebra, but several
Bayesian-focused texts have been published recently that are appropriate
electives for undergraduates with a calculus background
\citep{albert19, johnson21}. \endgroup

In many introductory courses, the only explicit connection to Bayesian
reasoning is Bayes' theorem taught as a formula derived from conditional
probabilities. Students are not invited to think in terms of prior and
posterior beliefs or coherent updating in the face of data. This
omission is increasingly at odds with contemporary practice. In AI and
machine learning, there has been a substantial increase in the
utilization of Bayesian reasoning, particularly in high-stakes domains
such as medical care and finance \citep{Ghahramani15, Chen21, Martin24}.
In addition, Bayesian procedures have shown advantages in predictive
performance and uncertainty calibration
\citep{Raftery97, Geweke06, Kendall17, Yu25}. There has also been a
noticeable increase in Bayesian research within academia across
disciplines, including psychology, neuroscience, and social sciences
\citep{Andrews13, Lynch19, Mueller24}. As a result, there is a widening
gap between the methods students see in their first statistics course
and the Bayesian approaches they encounter in advanced coursework,
research, and practice.

One reason for this gap is that fully fledged Bayesian inference and
computation require probability and mathematical machinery beyond the
scope of an introductory statistics course \citep{Dogucu22}. However,
recent work shows that Bayesian updating can be taught through active
learning in a single class session. \citet{Dogucu25} guide pre-service
mathematics and science teachers through prior expression and iterative
belief updating using an app-supported Beta-Binomial workflow.
\citet{Eadie19} use candy counts to motivate a Beta-Binomial model with
explicit prior and posterior interpretation.

Our proposed approach shares the same goal but makes different design
choices to fit the constraints of an introductory statistics course.
Rather than introducing an unknown parameter and a posterior
distribution, we frame Bayes' theorem through two-way tables and natural
frequencies that students can compute by hand. This emphasizes belief
updating as ordinary conditional probability over hypotheses as proposed
in \citet{Rossman95} and \citet{Hoegh20}, with minimal added technology
or distributional prerequisites. To achieve this, we developed an
activity in the style of Process Oriented Guided Inquiry Learning
(POGIL) that situates students in self-managed teams and guides them
through a sequence of carefully structured models and questions
\citep{POGIL25}.

The GAISE College Report recommends that introductory statistics courses
emphasize conceptual understanding, use real data with context and
purpose, and promote active learning environments in which students
learn by doing rather than by passively receiving information
\citep{GAISE16}. POGIL is an instructional approach aligned with these
recommendations. In a POGIL classroom, students work in small groups
with assigned roles, using structured activity sheets that lead them to
explore data, construct definitions, and derive relationships, while the
instructor acts as a facilitator. POGIL has been used successfully in
chemistry, biology, and other STEM (science, technology, engineering,
and mathematics) fields \citep{Minderhout07, Brown10, Kussmaul12}.
However, empirical evidence on the effectiveness of POGIL for teaching
Bayesian reasoning in introductory statistics is limited, and there are
few published activities that instructors can readily adopt.

\begingroup\color{black}The paper has two connected aims. First, we
describe a classroom-ready POGIL-style activity for teaching conditional
probability, Bayes' theorem, and belief updating in an introductory
statistics course. We focus on the activity itself, the educational
context in which it was implemented, and the design choices that make it
feasible in a course with limited time and minimal technical
prerequisites. \endgroup \begingroup\color{black} Second, we report a
quasi-experimental, between-cohort comparison in which the Bayes'
theorem unit was taught with either the POGIL-style activity or a
lecture-based approach. For this evaluation component, we examine
whether instructional condition was associated with (i) Bayes' theorem
final-exam performance and (ii) student satisfaction, while adjusting
for major field, gender, and race. To keep the emphasis on the
pedagogical contribution, the main text highlights the activity, the
study design, and the substantive findings, while full model
specifications and additional diagnostics are provided in the
supplementary material. Classroom materials referenced throughout the
paper are available through the supplementary material and the Open
Science Framework repository described in the Data Availability
Statement.\endgroup

Guided by these aims, we address the following questions:

\begin{enumerate}
\def\labelenumi{\arabic{enumi}.}
\tightlist
\item
  How do exam scores on Bayes' theorem content compare between students
  in POGIL and lecture sections?
\item
  How do satisfaction ratings with instruction compare between students
  in POGIL and lecture sections?
\item
  How are exam scores and satisfaction associated with students' major
  field, gender, and race?
\end{enumerate}

\section{Data and Context}\label{sec-data}

\subsection{Institutional Setting and Course
Context}\label{institutional-setting-and-course-context}

Data on student learning outcomes related to Bayes' theorem were
collected from undergraduate students enrolled in multiple sections of
an introductory statistics course, MAT 117 Elementary Statistics, at a
liberal arts college in the Midwest with approximately 1000
undergraduates. Most students in the course are freshmen, with smaller
numbers of sophomores, juniors, and seniors. Students typically enroll
in MAT 117 either because it is required for their major or to satisfy
the quantitative requirement for graduation.

\subsection{\texorpdfstring{\textcolor{black}{POGIL-style Activity and Instructional Placement}}{}}\label{section}

\begingroup\color{black}Before describing the between-cohort comparison,
we first describe the activity itself because it is the primary
pedagogical contribution of this paper. The Bayes' theorem lesson was
implemented as a self-contained unit within MAT 117 Elementary
Statistics, using hand-computable probabilities, two-way tables, and
structured group roles so that instructors could adopt the lesson
without specialized software or advanced probability prerequisites.
\endgroup \begingroup\color{black} Complete student and instructor
versions of the activity are provided in the supplementary material, and
the broader set of instructional materials, deidentified data, and
analysis code is archived in the Open Science Framework repository
described in the Data Availability Statement.\endgroup

\begingroup\color{black}The stand-alone nature of the activity allows it
to be used anytime in an introductory statistics course that works with
the instructor's planned course sequence. We indicate in the activity
facilitation notes that some previous introduction to probability may be
helpful, but the only prerequisite assumed is high school algebra. In
the fall and spring semesters where the POGIL-style activity and the
lecture teaching styles were compared, Bayes' theorem was introduced
toward the end of the semester after data collection, data description,
and frequentist inference were covered.\endgroup

\begingroup\color{black} The introduction to the POGIL-style activity or
the lecture on Bayes' theorem included an overview of conditional
probabilities and associated notation as well as an explanation of the
difference between a frequentist approach and Bayesian reasoning in
which the data update prior probabilities. We explained to students the
advantage of being able to calculate a probability for the null and the
alternative hypotheses in a Bayesian approach instead of needing
dichotomous decision making about the null hypothesis in the frequentist
approach. Instructors implementing the activity could use it as a
springboard for more Bayesian reasoning, or leave it as a stand-alone
introduction.\endgroup

The POGIL-style activity was developed through participation in the
Bayes-BATS instructor training program.\footnote{\url{https://bayes-bats.ics.uci.edu/}}
One author attended a week-long summer workshop (Tier 1) in which
Bayesian reasoning was taught through lecture and hands-on activities.
She and two other participants subsequently proposed and received
support for a Tier 2 project to develop up to three POGIL-style
activities that would require students to (1) develop Bayes' theorem,
(2) apply Bayes' theorem, and (3) use the Beta-Binomial model in
Bayesian reasoning. Two of these activities were fully developed and
piloted in several classrooms.

\begin{figure}

\centering{

\includegraphics[width=4.5in,height=\textheight,keepaspectratio]{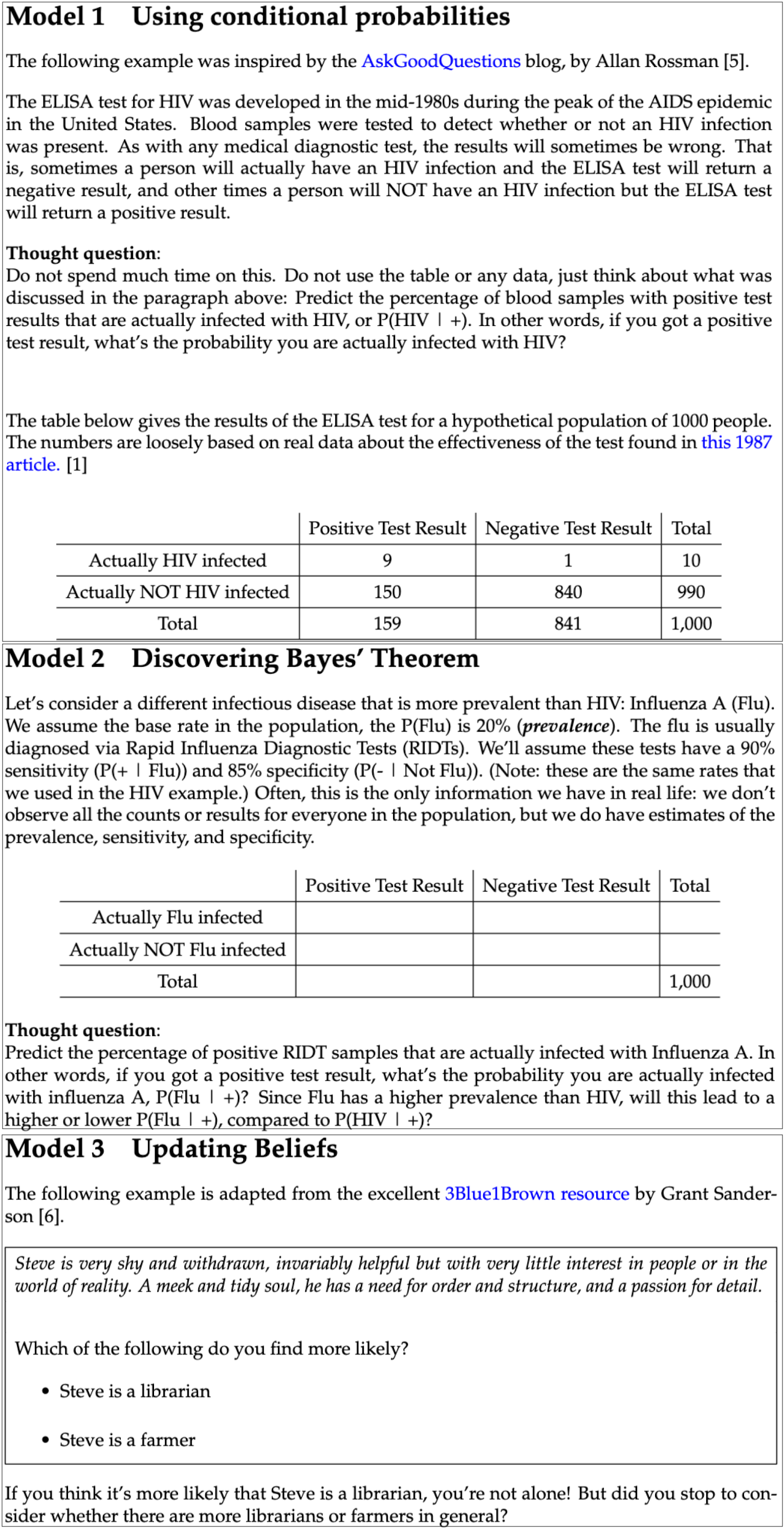}

}

\caption{\label{fig-pogil-model}Screenshot excerpts from the three
models in the POGIL-style activity.}

\end{figure}

As part of a Northwestern Mutual Data Science Institute (NMDSI) student
scholar award, the first two activities were redesigned and combined
into a single, longer activity that served as the focus of the present
study. The combined POGIL-style activity comprised 14 pages organized
into three models (Figure~\ref{fig-pogil-model}). Model 1, \emph{Using
Conditional Probabilities}, leads students through the analysis of a
two-way table with fabricated ELISA (enzyme-linked immunosorbent assay)
screening test results, positive or negative, and infection status, HIV
infected or not infected, inspired by Allan Rossman's AskGoodQuestions
blog \citep{Rossman19}. Students compute and interpret conditional
probabilities in this context. Model 2, \emph{Discovering Bayes'
Theorem}, presents an empty two-way table that students fill in with
Influenza A infection status and RIDT (Rapid Influenza Diagnostic Tests)
test results. As they work through the relationships among table cells,
students effectively ``discover'' Bayes' theorem. Model 3,
\emph{Updating Beliefs}, requires students to apply Bayes' theorem to a
non-medical scenario adapted from an example involving a description
that might match a farmer or a librarian more closely
\citep{kahneman11}. Across models, the activity aligns with GAISE
College Report recommendations by emphasizing conceptual understanding,
active student engagement, and real or realistic contexts.

\subsection{Study Design and Instructional
Conditions}\label{study-design-and-instructional-conditions}

The study used a quasi-experimental, between-cohort design implemented
in back-to-back fall and spring semesters. During Fall 2024, two
sections of MAT 117 taught by two different professors introduced Bayes'
theorem using the POGIL-style activity. During Spring 2025, two
additional sections of the same course were taught by the same two
professors, but Bayes' theorem was covered using a traditional
lecture-based approach. Each section enrolled approximately 30 students,
and each professor taught in the same physical classroom in both
semesters.

One professor holds a PhD in mathematics and teaches in the mathematics
program; the other holds a PhD in soil science and has 18 credits of
graduate level statistics. Because instructor effects are a common
alternative explanation in quasi-experimental comparisons, the
implementation was designed to reduce systematic differences
attributable to instructor identity and to make the Bayes' theorem
instructional approach the primary planned difference between
conditions. In particular, both professors used the same course
structure, learning objectives, and course materials across semesters,
with the single planned exception of the delivery mode for the Bayes'
theorem unit. \begingroup\color{black} Additional details and
explanation are provided in the supplementary material.\endgroup

In the POGIL semester, the Bayes' theorem materials consisted of brief
introductory slides followed by the POGIL-style activity described
above. In the lecture semester, the Bayes' theorem materials consisted
of introductory slides, lecture slides, and an in-class worksheet. To
ensure that any observed differences were not driven by unequal practice
opportunities or assessment content, the out-of-class worksheet and the
final exam questions on Bayes' theorem were identical across all
sections in both semesters. The student demographic survey was also
identical across sections and semesters, while the satisfaction survey
items were necessarily phrased to reflect the instructional approach,
POGIL or lecture, so that students were evaluating the mode of
instruction they experienced. Finally, student course evaluations did
not indicate a systematic preference for one instructor over the other,
providing additional contextual information about possible instructor
differences.

\begin{figure}

\centering{

\includegraphics[width=1\linewidth,height=\textheight,keepaspectratio]{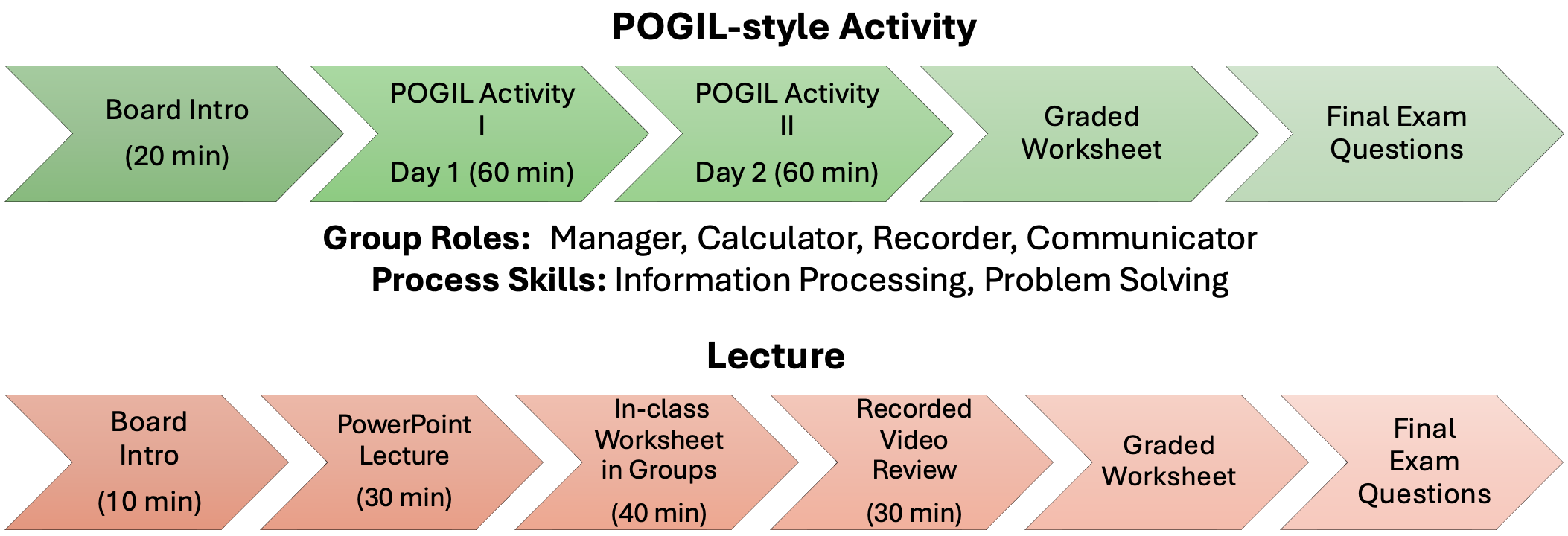}

}

\caption{\label{fig-pogil-flow}Flow chart comparing the two pedagogical
approaches used in this study.}

\end{figure}

In the Fall 2024 semester, both professors used two 80-minute class
meetings to complete the POGIL-based instruction on Bayes' theorem (see
Figure~\ref{fig-pogil-flow}). On the first day, class began with a brief
review of conditional proportions and probabilities led by the professor
at the board. A short slide-based activity then asked students to
respond to questions that placed them on a frequentist and Bayesian
continuum, adapted from an idea in \emph{Bayes Rules!}
\citep{johnson21}. After about 20 minutes, students were organized into
groups of four, desks were rearranged so that group members could face
one another, and the POGIL-style activity packets were distributed. Each
student assumed a specific POGIL role including manager, calculator,
recorder, and communicator. Groups worked at their own pace through the
activity, while the professor circulated to answer questions and
facilitate discussion. In the next class meeting, the same groups
reformed and completed the activity. The time required to complete the
activity varied across groups, with some finishing early and others
requiring additional time; on average, students spent approximately 60
minutes working on the activity on the first day, and another 60 minutes
working on it the second day. At the end of the second class, students
received an out-of-class worksheet on Bayes' theorem to complete for a
grade, which also served as preparation for the final exam questions.

In the Spring 2025 semester, both professors used one 80-minute class
session to cover Bayes' theorem using a lecture-based approach (see
Figure~\ref{fig-pogil-flow}). Class began with a 10-minute review of
conditional proportions, followed by a 30-minute lecture using slides to
introduce and work through examples of Bayes' theorem. Students were
then given an in-class worksheet to complete in small groups to practice
applying Bayes' theorem (see the supplementary material and the OSF for
descriptions and full copies of the presentation slides and in-class
worksheet). At the end of class, students received the same out-of-class
worksheet that had been used in the fall semester to complete for a
grade. In the spring, students also had access to a recorded video in
which the instructor worked through the in-class worksheet. The total
instructional time devoted to Bayes' theorem, including in-class and
optional out-of-class resources, \begingroup\color{black}was 110 minutes
for the lecture semester and on average 140 minutes for the
POGIL-activity semester. \endgroup In both semesters, the out-of-class
worksheet, which was collected for a grade, prepared students for the
final exam questions on Bayes' theorem administered during the
cumulative final exam in finals week. Students also completed a
demographic and a satisfaction survey at the end of the semester (see
the supplementary material and the OSF for descriptions and full copies
of these surveys).

\subsection{Variables and Measures}\label{variables-and-measures}

\begingroup\color{black}For the evaluation component of the study, the
analytic dataset consisted of student survey responses and Bayes'
theorem final-exam question scores linked at the individual level. The
variables below were screened for completeness, recoded when necessary,
and transformed to support consistent interpretation and stable
estimation with a moderate sample size.\endgroup

\begingroup\color{black}\textbf{Outcome variables.} The primary outcomes
were Bayes' theorem final-exam performance and instructional
satisfaction. The final exam performance (\texttt{finalexam}) ranged
from 0 to 100 points and was based on student scores from the Bayes'
theorem questions on the cumulative final exam. Scores were divided by
100 to produce a variable on the unit interval {[}0, 1{]} suitable for
modeling with a zero--one inflated Beta distribution. Approximately one
quarter of students earned perfect scores, yielding a pronounced ceiling
effect that justified a non-Gaussian specification. The satisfaction
outcome was recorded originally on a 1--10 Likert-type scale (1 = very
unsatisfied, 10 = very satisfied). Raw distributions of these ordered
categories and a sensitivity analysis using the original 1--10 response
are reported in the supplementary material. For the POGIL group, the raw
question asked about satisfaction with the group work aspect of the
activity; for the lecture group, it asked about satisfaction with the
lecture and slides. For the main ordinal model, these raw responses were
recoded into \texttt{satis\_order}, an ordered factor with three levels,
Low (1--4), Medium (5--7), and High (8--10), to reflect the ordinal
nature of the data and to stabilize estimation with limited sample
size.\endgroup

\begingroup\color{black}\textbf{Explanatory variables.} The primary
explanatory variable was instructional style (\texttt{teachstyle}),
indicating whether Bayes' theorem was taught using a POGIL-style
activity or a traditional lecture. Students' major type (\texttt{stem})
was categorized as STEM, non-STEM, or undecided. Gender
(\texttt{gender}) was self-reported and treated as a categorical
variable. The survey also included ``Non-binary'' and ``Prefer not to
say'' options, but in this sample all students selected either Male or
Female. Race (\texttt{race}) was initially reported in five categories:
Asian, Black, White, Two or more, and Other. Because several categories
contained fewer than three students, these were collapsed into White and
Other Racial Identities (ORI) groups to produce stable posterior
estimates. These covariates were included to describe the sample and to
support adjusted comparisons across instructional conditions.\endgroup

\textbf{Data summary.} \begingroup\color{black}Data cleaning involved
matching student outcomes (e.g.~final-exam score) with survey responses,
removing identifying information, and coding as needed for consistency.
Students who did not complete one or both surveys or who submitted
incomplete surveys were removed from the dataset. \endgroup After
cleaning, the final analytic dataset contained 94 complete student
records. Exam scores were concentrated near 100, and satisfaction
ratings were concentrated in the High category. These features motivated
the use of outcome-specific Bayesian models in the evaluation stage of
the paper.

\section{Statistical Methods and Analysis}\label{sec-method}

\begingroup\color{black}The goal of the analysis was to examine whether
instructional style (POGIL vs lecture) was associated with students'
Bayes' theorem exam performance and instructional satisfaction while
accounting for major type, gender, and race. Because the two outcome
variables differed in scale and distribution, we used a Bayesian
bivariate generalized linear modeling framework with conditionally
independent outcome components, matching each outcome to an appropriate
likelihood. The main text reports only the essential modeling rationale
needed to interpret the results; full model specifications, prior
distributions, and additional diagnostic details are provided in the
supplementary material.\endgroup

\subsection{Bayesian Bivariate Generalized Linear
Model}\label{bayesian-bivariate-generalized-linear-model}

\begingroup\color{black} The model included two conditionally
independent outcome components. Bayes' theorem exam scores were modeled
with a zero--one inflated Beta distribution because the scores were
bounded between 0 and 100 and included many perfect scores. Satisfaction
was modeled with a cumulative logit formulation because the recoded
response was ordinal with three levels: Low, Medium, and High. Both
submodels used the same predictors: instructional style, major type,
gender, and race. We also screened a small set of
teaching-style-by-demographic interaction terms, but the final reported
model retained only main effects because the interaction models did not
improve out-of-sample predictive performance. Additional equations,
model-comparison details, and diagnostic results are provided in the
supplementary material.\endgroup

\subsection{Prior Distributions and Posterior
Inference}\label{prior-distributions-and-posterior-inference}

\begingroup\color{black}We used weakly regularizing priors and fit the
model in \texttt{brms} with Stan. Posterior summaries are reported as
means and 95 percent credible intervals, and model-based predictions are
translated back to exam points and satisfaction probabilities for
interpretation. Full prior specifications, sampling settings,
convergence diagnostics, and posterior predictive checks are provided in
the supplementary material.\endgroup

\section{Results}\label{sec-results}

\subsection{Descriptive Findings}\label{descriptive-findings}

\begin{table}

\caption{\label{tbl-demo}Participant demographics by instructional
condition.}

\centering{

\fontsize{9.0pt}{10.8pt}\selectfont
\begin{tabular*}{450pt}{@{\extracolsep{\fill}}lcc}
\toprule
\textbf{Variable} & \textbf{Lecture}  N = 46 & \textbf{POGIL}  N = 48 \\ 
\midrule\addlinespace[2.5pt]
{\bfseries Gender} &  &  \\ 
    Female & 28 (61\%) & 25 (52\%) \\ 
    Male & 18 (39\%) & 23 (48\%) \\ 
{\bfseries Race} &  &  \\ 
    Other Racial Identities & 4 (8.7\%) & 5 (10\%) \\ 
    White & 42 (91\%) & 43 (90\%) \\ 
{\bfseries Major Type} &  &  \\ 
    Non-STEM & 19 (41\%) & 15 (31\%) \\ 
    STEM & 25 (54\%) & 27 (56\%) \\ 
    Undecided & 2 (4.3\%) & 6 (13\%) \\ 
\bottomrule
\end{tabular*}

}

\end{table}%

A total of 94 students contributed complete data: 48 in the POGIL
sections and 46 in the lecture sections. Table~\ref{tbl-demo} summarizes
demographic characteristics by instructional condition. The two groups
were similar in gender and race, with a slightly higher proportion of
non-STEM majors in the lecture group. Most students were first-year
undergraduates taking the course to satisfy a general-education
requirement.

\begin{figure}

\centering{

\includegraphics[width=1\linewidth,height=\textheight,keepaspectratio]{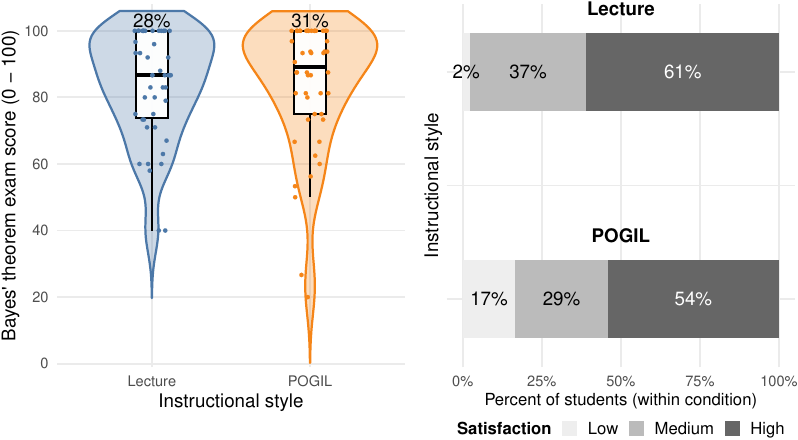}

}

\caption{\label{fig-dist}Distributions of the response variables by
instructional condition. Left: Bayes' theorem exam scores (0 -- 100)
shown with violin plots, boxplots, and individual student scores. The
percentages above the groups indicate the proportion of students
obtaining perfect scores. Right: Satisfaction ratings
(Low/Medium/High).}

\end{figure}%

Figure~\ref{fig-dist} displays the distribution of Bayes' theorem exam
scores and the ordinal satisfaction outcome by instructional condition.
In the left panel, the violin plots, boxplots, and individual student
scores show substantial overlap between the Lecture and POGIL score
distributions, with similar medians in the mid-80s and a visible
concentration near 100. Perfect scores occurred for 13 Lecture students
(28\%) and 15 POGIL students (31\%). The violin plots with observations
shown make the ceiling effect easy to see directly, and the boxplots
highlight the similarity of the central score distributions across
instructional conditions. Ratings were skewed toward the High category
for both instructional formats, indicating that students were generally
satisfied regardless of delivery mode. Only a small fraction of
responses fell in the Low range. These patterns motivated the use of an
ordinal cumulative-logit model to account for ordered yet unevenly
spaced categories.

\begingroup\color{black}The supplementary material displays the original
1--10 satisfaction responses as ordered categories and reports
sensitivity analyses based on the raw ordered response and two
alternative regroupings. Across these codings, the estimated
teaching-style contrast in satisfaction remains directionally
consistent, with POGIL generally lower than Lecture on upper-tail
satisfaction probabilities, but its magnitude depends on how the upper
end of the scale is defined.\endgroup

\subsection{Model-Based Results}\label{model-based-results}

\begin{figure}

\centering{

\includegraphics[width=1\linewidth,height=\textheight,keepaspectratio]{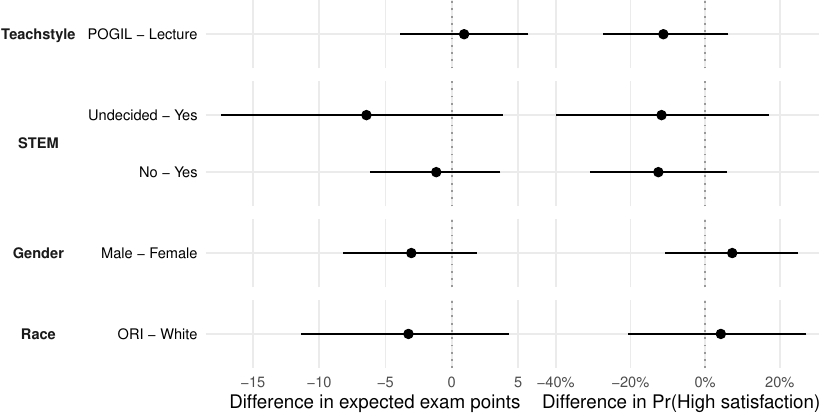}

}

\caption{\label{fig-forest}95\% credible intervals for model-based
posterior differences from the retained main-effects model. The black
dots show posterior mean differences after averaging predictions over
the observed distribution of the remaining covariates. ORI stands for
Other Racial Identities. Left: expected exam points. Right: probability
of High satisfaction.}

\end{figure}

\begingroup\color{black}The summaries in Table~\ref{tbl-effect} and the
contrasts in Figure~\ref{fig-forest} come from the retained main-effects
model (M0). In plain language, for a given comparison we asked what the
fitted model would predict if all students in the analytic sample were
assigned, in turn, to each level of that factor while keeping their
other observed characteristics the same, and then averaged those
model-based predictions across the sample. This yields
covariate-adjusted posterior summaries without changing the substantive
conclusions of the analysis. Accordingly, the expected exam score refers
to the full posterior mean of the zero--one inflated Beta outcome on the
original 0--100 scale, including the model-implied point mass at perfect
scores rather than only the continuous component. We retain this
expectation-based summary in the main text because the zero--one
inflated Beta model is designed to describe exactly the kind of score
distribution observed here: after rescaling scores to the unit interval,
the Beta component flexibly accommodates skewness among values between 0
and 1, while the inflation component separately captures the extra
pile-up at perfect scores. When transformed back to the original 0--100
scale, the fitted model provides a good description of the observed
score distribution, as shown by the posterior predictive checks in the
supplementary material. For completeness, the supplementary material
also reports posterior predictive medians by instructional style. Those
medians are likewise very similar across instructional conditions and
lead to the same practical conclusion, but we treat them as a secondary
descriptive check because the fitted model is parameterized in terms of
the full expectation of the outcome.\endgroup

Table~\ref{tbl-effect} summarizes expected exam performance and the
probability of reporting High satisfaction by teaching style, STEM major
status, gender, and race. Expected scores were similar across groups,
with posterior means in the low-to-mid 80s and wide 95 percent credible
intervals. Posterior means for High satisfaction were also comparable
across factors, and interval overlap was substantial, indicating
considerable uncertainty.

\begingroup\color{black}Supplementary sensitivity analyses using the
original ordered 1--10 responses and alternative regroupings yielded the
same directional pattern for satisfaction, with POGIL generally lower
than Lecture on upper-tail satisfaction probabilities, but the magnitude
of the contrast depended on the chosen threshold. In particular, broader
upper-tail definitions such as 7--10 or 6--10 produced more negative
contrasts than the pre-specified main comparison based on High =
8--10.\endgroup

\begin{table}

\caption{\label{tbl-effect}Model-based posterior summaries from the
retained main-effects model. Each estimate is obtained by averaging
predictions over the observed distribution of the remaining covariates
in the analytic sample. The expression \(x\) \([y, z]\) means posterior
mean \(x\) with 95\% credible interval \([y, z]\).}

\centering{

\fontsize{9.0pt}{10.8pt}\selectfont
\begin{tabular*}{\linewidth}{@{\extracolsep{\fill}}llll}
\toprule
{\bfseries Effect} & {\bfseries Level} & {\bfseries Exam scores} & {\bfseries Pr(High satisfaction)} \\ 
\midrule\addlinespace[2.5pt]
Teachstyle & Lecture & 82.1 [77.2, 86.4] & 62.5\% [49.4\%, 74.7\%] \\ 
Teachstyle & POGIL & 83.0 [78.4, 87.1] & 51.4\% [37.9\%, 65.0\%] \\ 
STEM & Yes & 83.5 [78.9, 87.6] & 62.3\% [49.4\%, 74.2\%] \\ 
STEM & No & 82.3 [77.4, 86.7] & 49.8\% [35.3\%, 64.9\%] \\ 
STEM & Undecided & 77.1 [66.3, 86.5] & 50.7\% [25.3\%, 75.9\%] \\ 
Gender & Female & 83.9 [79.6, 87.7] & 53.7\% [41.1\%, 66.4\%] \\ 
Gender & Male & 80.8 [75.5, 85.5] & 60.9\% [46.3\%, 74.1\%] \\ 
Race & White & 82.9 [78.7, 86.5] & 56.4\% [46.1\%, 66.4\%] \\ 
Race & Other Racial Identities & 79.6 [71.2, 86.9] & 60.7\% [37.4\%, 82.2\%] \\ 
\bottomrule
\end{tabular*}

}

\end{table}%

Figure~\ref{fig-forest} visualizes the model-based posterior differences
in expected exam points and in the probability of High satisfaction. For
teaching style, the posterior mean difference POGIL minus Lecture in
expected exam points was small, and the 95 percent credible interval
included 0. The posterior difference in the probability of High
satisfaction was likewise small with a 95 percent credible interval that
spanned 0. For STEM status, contrasts for no versus yes and undecided
versus yes showed modest differences in both outcomes with wide
intervals that included 0. For gender and race, posterior adjusted
differences in expected points and in the probability of High
satisfaction were also centered near 0 with intervals that overlapped 0.
Overall, the model-based results suggest no clear evidence of systematic
differences in either exam performance or High satisfaction across the
studied factors in this sample.

The model also yields a direct estimate of the probability of a perfect
exam score. Differences in Pr(score = 100) by teaching style and
demographic groups were small with wide intervals; we report these in
the supplementary material for completeness.

\subsection{\texorpdfstring{\textcolor{black}{Model Adequacy Checks}}{}}\label{section-1}

\begingroup\color{black}We assessed model adequacy using posterior
predictive checks, but because these checks serve mainly as technical
validation rather than as the primary pedagogical contribution of the
paper, we now summarize them briefly here and report the figures and
fuller discussion in the supplementary material.\endgroup

\begingroup\color{black}Overall, the fitted model reproduced the main
features of the observed data, including the concentration of Bayes'
theorem exam scores near 100 and the strong skew of satisfaction ratings
toward the High category across instructional conditions. These checks
support the adequacy of the modeling framework used for the adjusted
comparisons reported above.\endgroup

\subsection{Interpretation and Educational
Relevance}\label{interpretation-and-educational-relevance}

Although the posterior contrasts for teaching style, major type, gender,
and race were centered near zero with wide credible intervals, the
results are still informative. Taken together, they suggest that
students in Lecture and POGIL sections achieved comparable final exam
performance and reported similar satisfaction. The Bayesian framework
supports a probabilistic interpretation rather than a binary one. The
data do not provide strong evidence favoring either instructional
approach, while still allowing for the possibility of modest differences
that a larger or more targeted sample could detect.

\begingroup\color{black}One possible explanation for the similarity
between instructional styles is that the lecture condition in this study
was not a purely passive lecture format. Both instructors regularly used
hands-on group activities and problem-based practice throughout MAT 117,
so students in both semesters were accustomed to active engagement
during class. Even in the lecture sections for the Bayes' theorem unit,
students received direct instruction followed by group work on the
in-class worksheet. As a result, the lecture condition may have included
more active learning than is typically implied by the term ``lecture.''
This may have reduced the contrast between the two instructional
conditions and made large differences in exam performance or
satisfaction less likely to appear.\endgroup

From a pedagogical perspective, these findings indicate that
implementing an active learning approach such as POGIL need not come at
the expense of student outcomes. In this setting, collaborative,
inquiry-based instruction performed on par with traditional lecture,
which supports the view that multiple instructional formats can be
effective in introductory statistics and gives instructors greater
flexibility to select strategies that fit their context, constraints,
and teaching goals.

More broadly, prior research indicates that activity-based teaching does
not consistently improve exam performance
\citep{Dochy03, Setren21, BuhlWiggers23}. Meta-analyses in undergraduate
STEM report average gains under active learning but also substantial
variability, with some implementations producing little improvement or
lower exam scores \citep{Freeman14, Gillette18}. In statistics
education, effects appear sensitive to how activities are structured.
Studies that emphasize collaborative small group work often report
positive achievement gains, whereas studies of inquiry-based formats
often report average effects that are small or statistically
indistinguishable from zero \citep{Kalaian14, Mesghina24}.

Results from problem-based learning and flipped classrooms likewise
range from positive to null, plausibly because insufficient scaffolding
increases cognitive load and weakens performance on conventional
assessments \citep{Kirschner06}.

Methodologically, the analysis illustrates how Bayesian modeling
supports responsible interpretation of educational data. By jointly
modeling performance and satisfaction with realistic outcome
distributions, the study moves beyond conventional significance testing
to emphasize model adequacy, uncertainty quantification, and effect
plausibility. This approach models the kind of reasoning educators can
adopt when evaluating innovations in their own classrooms: focus on the
credibility of evidence and its practical meaning, not just the presence
or absence of statistical significance.

\section{Discussion}\label{sec-disc}

This study had two complementary goals. The first was to introduce a
process-oriented guided-inquiry learning approach for teaching Bayesian
reasoning in introductory statistics. To our knowledge, this is among
the first published implementations of a POGIL-style activity
specifically designed to guide students through Bayesian concepts such
as conditional probability, prior and posterior reasoning, and the logic
of evidence updating. We developed, piloted, and refined a sequence of
models that encourage students to construct Bayesian ideas
collaboratively rather than receive them passively. This POGIL-style
activity, provided in the supplementary material, may serve as a
starting point for instructors who wish to adapt inquiry-based learning
to other areas of statistical reasoning.

\begingroup\color{black} Our activity is closely related in spirit to
Albert and Rossman's \textit{Workshop Statistics} approach, which
minimizes lecture time, emphasizes students working actively with data,
and introduces basic statistical inference from a Bayesian viewpoint
\citep{albert01}. The main difference is one of instructional form and
scope. \textit{Workshop Statistics} provides a broader activity-based
curriculum, whereas our contribution is a modular POGIL-style lesson
that can be inserted into a conventional introductory statistics course.
In addition, the POGIL format explicitly incorporates structured team
roles, guided inquiry, and process-skill development, which are central
features of POGIL.\endgroup

\begingroup\color{black} The disease-testing context in our Model 1 is
similar to the context used in Albert and Rossman's in-class activity
15-1, ``Do you have a rare disease?'' \citep{albert01}. We adapted this
context into a POGIL-style sequence in which students first use
conditional probabilities in a completed two-way table, then develop
Bayes' theorem through a second disease-testing example, and finally
apply Bayesian updating in a non-medical context. Thus, our activity
extends prior Bayesian activity-based work by organizing Bayes' theorem
and belief updating into a role-based guided-inquiry structure suitable
for a single instructional unit.\endgroup

The second goal was to examine, using a transparent Bayesian modeling
framework, how students in POGIL and Lecture sections compared in exam
performance and satisfaction. The analysis incorporated realistic
outcome distributions, a zero--one inflated Beta model for exam scores
and an ordinal logit model for satisfaction, and used these models to
estimate adjusted comparisons with uncertainty. Across both outcomes,
the posterior contrasts were centered near zero with wide credible
intervals, indicating no clear evidence of systematic differences
between the two instructional conditions. \begingroup\color{black}
Because instructional style was implemented by semester rather than
randomly assigned within semester, these model-based comparisons should
be interpreted as covariate-adjusted associations in this
quasi-experimental setting rather than as definitive causal effects.
\endgroup These results also illustrate the value of Bayesian inference
for educational research: the analysis quantifies uncertainty instead of
dichotomizing results and demonstrates that similar performance and
satisfaction are plausible under both teaching styles.

\begingroup\color{black} The scope of inference for this study is
necessarily limited. The findings are based on four sections of a single
introductory statistics course at one institution during one academic
year and are therefore most directly relevant to similar instructional
settings, rather than to all introductory statistics courses,
institutions, or implementations of POGIL-based Bayesian instruction. In
addition, the demographic composition of the small liberal arts college
where the study was conducted resulted in limited variation in gender
and race, so the analysis necessarily simplifies a more complex reality.
Future work should examine these questions in broader and more diverse
student populations.\endgroup

\begingroup\color{black}The outcome measures also have important
limitations. The Bayes' theorem exam score reflects performance on a
small set of course assessment items and therefore provides evidence
about short-term course performance rather than longer-term retention or
transfer of Bayesian reasoning. The satisfaction outcome should also be
interpreted cautiously because the survey items were parallel but not
identical across conditions: students in the POGIL sections rated
satisfaction with the group-work aspect of the activity, whereas
students in the lecture sections rated satisfaction with the lecture and
slides. As a result, the satisfaction comparison is informative about
students' reported instructional experience, but it should not be
interpreted as a precise measure of the comparative effectiveness of the
two approaches. Supplementary sensitivity analyses further show that the
estimated teaching-style contrast in satisfaction is directionally
stable across several ordinal codings, but not numerically identical;
stronger contrasts appear when broader upper-tail thresholds are used.
This reinforces both the importance of respecting the ordinal nature of
the response and the need to avoid over-interpreting any single cutpoint
choice.\endgroup

These findings should be interpreted in context. The data come from the
instructors' first implementation of a POGIL-style activity in this
course. As familiarity with guided-inquiry facilitation grows, learning
gains and satisfaction may evolve. Ongoing data collection in future
course offerings would allow longitudinal modeling to distinguish
instructional effects from instructor learning effects. Thus, the
present analysis should be viewed as a baseline in a potential multiyear
investigation of POGIL's impact on students' Bayesian reasoning.

Finally, this study emphasizes full transparency and reproducibility.
All deidentified data, analysis code, and instructional materials
supporting the study are archived in the Open Science Framework
repository listed in the Data Availability Statement, and the
supplementary material provides direct guidance on how those resources
connect to the manuscript. Although the conclusions are not definitive
due to modest sample size and large posterior uncertainty, the study
demonstrates how transparent, reproducible Bayesian analyses can inform
teaching practice responsibly. The combination of pedagogical innovation
and rigorous modeling provides a framework for future work that both
deepens our understanding of Bayesian reasoning in students and
exemplifies the evidence-based approach that statistics education seeks
to promote.

\section{Acknowledgments}\label{sec-ackn}

The authors gratefully acknowledge Katie Fitzgerald and Olga Glebova for
their contributions to the ideas and the initial formulation of the
models and the first drafts of the POGIL-style activities. They also
participated in piloting these drafts in their own classrooms. We also
acknowledge Monika Hu, Mine Dogucu, and Amy Herring for their work as
Principal Investigators for the Bayes-BATS instructor training bootcamp
one author attended as part of the Tier 1 aspect of the Bayes-BATS
program, as well as their support for developing Bayesian teaching and
learning materials that resulted in the first drafts of the POGIL-style
activities. Additionally, the authors gratefully thank Tova Brown for
her efforts in this project, both for her willingness and enthusiasm to
teach Bayes' theorem using the POGIL-style activity and the lecture
approach and for her commitment to collecting data from the students in
her sections of MAT 117.

\section{Funding}\label{funding}

This project was made possible in part by travel support from
Bayes-BATS, funded by the National Science Foundation Improving
Undergraduate STEM Education Program (NSF IUSE: EHR; award numbers
2215879, 2215920, and 2215709), which supported attendance at the
week-long Bayes-BATS workshop. Additional funds from the same
NSF-supported project supported work on the initial drafts of the
POGIL-style activities. The Northwestern Mutual Data Science Institute
(NMDSI) also awarded the authors a grant to support the initial
classroom data collection process.

\section{Disclosure Statement}\label{disclosure-statement}

The authors report no potential conflicts of interest. The study was
reviewed and approved by the Institutional Review Board (IRB) at the
authors' university.

During the preparation of this manuscript, the authors used ChatGPT
(OpenAI, GPT-5.5, accessed June 9, 2026) to assist with language
editing, paraphrasing, and coding assistance. All AI-generated outputs
were reviewed, edited, and verified by the authors, who take full
responsibility for the content of the manuscript.

\section{Data Availability Statement}\label{data-availability-statement}

All deidentified data, analysis code, and instructional materials
supporting this study are available in an Open Science Framework
repository at: \url{https://tinyurl.com/yshb6wer}. The repository
includes README files with instructions for reproducing the analyses and
using the instructional materials.

\bibliography{bibliography.bib}

\end{document}